\documentclass[a4paper]{iopart}
\usepackage{graphicx}

\def\jpb{{\em J. Phys. B: At. Mol. Opt. Phys.}~}
\def\pra{{\em Phys. Rev. A}~}

\def\prl{{\em Phys. Rev. Lett.}~}
\def\jmo{{\em J. Mod. Opt.}~}
\def\rmp{{\em Rev. Mod. Phys.}~}

\def\jetp{{\em J. Exp. Theor. Phys.}~}
\def\etal{{\em et al. }}

\newcommand{\vecr}{\mathbf{r}}

\newcommand{\beq}{\begin{equation}}
\newcommand{\eeq}{\end{equation}}

\begin{document}
\title{Diagnostics of ultra-intense laser pulses using tunneling ionization}

\author{M. F. Ciappina}

\address{Institute of Physics of the ASCR, ELI-Beamlines project, Na Slovance 2, 182 21 Prague, Czech Republic}

\author{S. V. Popruzhenko}

\address{Prokhorov General Physics Institute RAS, Vavilova str.38, 119991 Moscow, Russia}
\address{Voronezh State University, Universitetskaya pl.1, 394036 Voronezh, Russia}

\ead{sergey.popruzhenko@gmail.com}

\begin{abstract}
We revisit a recently proposed scheme [M.F. Ciappina {\em et al} 2019 \pra {\bf 99} 043405] for accurate measurement of electromagnetic radiation intensities in a focus of high-power laser beams.
The method is based on the observation of multiple sequential tunneling ionization of atoms and suggests determination of the peak intensity value from the appearance in the focal area ions with a certain charge number.
Here we study the sensitivity of the approach to two essential factors: the focal volume effect and the tunneling ionization model chosen to calculate ionization rates.
We show, on the one hand, that the results appear almost model independent for ionization of helium- and hydrogen-like ions and, on the other hand, that the focal averaging leaves the intensity-dependent features visible.
Our findings are in favor to a practical implementation of the atomic diagnostics of extreme laser intensities. 
\end{abstract}

\section{Introduction}

The recent commissioning of several laser facilities of multi-petawatt (PW) power, including APRI \cite{APRI}, CAEP \cite{CAEP} and SULF \cite{SULF}, as well as the decisive progress in the construction of the ELI laser pillars~\cite{ELI,P3-ELI}, belonging to the same multi-PW class, open the door to laboratory studies of the interaction between extremely intense light and matter in the new, so far unexplored regimes.
The transition from sub-PW to multi-PW laser beams implies the extension of the experimentally available peak laser intensity from the currently accessible range $10^{20}\div 10^{21}{\rm W/cm}^2$ to $10^{22}\div 10^{23}{\rm W/cm}^2$.
Furthermore, the currently developing projects of even more powerful, sub-exawatt lasers sources \cite{XCELS}, promise reaching intensities up to $10^{25}\div 10^{26}{\rm W/cm}^2$.
These technical achievements will allow, in particular, to study in laser laboratories a number of new phenomena of classical and quantum physics, including radiation-dominated dynamics of isolated charges and plasmas, laser initiation of cascades of elementary particles, excitation of extremely strong magnetic fields, production of electron-positron pairs, etc. 
The respective research fields and the current achievements received a detailed description in a set of review articles~\cite{bulanov-rmp09,dipiazza-rmp12,fedotov-cp15}.

In order to provide a reliable correspondence between experimental data on the interaction of ultra intense laser radiation with matter and theoretical predictions, a detailed characterization of this radiation is required.
This includes, in particular, knowing the peak laser intensity value in the focus with reasonable accuracy.
For laser-induced and assisted phenomena, which may proceed in the strong-field regime, the dependence of observables on the laser intensity appears to be highly nonlinear so that a relatively small variation in its peak value may cause a considerable change in the response of the laser-irradiated target.
Thus, an accurate determination of the focal intensity distribution and the intensity peak value is crucially important to correctly interpret experimental results and connect them to theory.
At ultrahigh intensities, no direct measurement of the intensity distribution in the focus is possible with cameras or other similar detectors.
Extrapolation of measurements made in the low-power mode to the high-power regime is questionable because of the various nonlinear effects happening throughout the amplifying and focusing systems.
Therefore, the only reliable methods for characterization of the laser focus at extreme intensities can be those based on the observation of effects of laser-matter interactions sensitive to the laser parameters of interest. 

As far as the measurement of intensity is concerned, several essential constraints limit possible interactions scenarios. 
Firstly, the interaction must not make any significant effect on the electromagnetic field distribution in the focus.
This immediately excludes dense plasma and solids as possible targets.
For this reason only low-density atomic gases or particle beams can be used.
Secondly, the response should be sensitive more to the local value of intensity than to its global distribution in space and time.
Recently, several effects have been discussed and numerically examined as possible tools for the determination of laser peak intensities in the ultrahigh-power regime.
In Refs.~\cite{ciappina-pra19,ciappina-book19} sequential tunneling ionization of multielectron atoms has been considered.
It was shown that, owing to the highly nonlinear dependence of the tunneling ionization rate on the electric field strength of the laser wave, the maximal charge state of a given atomic specie produced in the laser focus appears highly sensitive to the peak value of intensity.
The same value is much less sensitive to the focal distribution of intensity and to the pulse duration.
This method roots back to the experimental work \cite{walker-pra01,walker-josa03,yamakawa-pra03,yamakawa-jmo03} where the commonly known analytic formulas for the rate of tunneling ionization \cite{ppt-jetp66a,pp-jetp67,popov-usp04} have been quantitatively verified by observing tunneling ionization of rare gases at intensities $\simeq 10^{19}{\rm W/cm}^2$.
Another proposal, based on the observation of an intensity-dependent shift in spectra of Thomson scattering, has been experimentally tested in~\cite{he-oe19}. 
In this approach, the wavelength shifts proportional to the laser intensity, appear to be in fair agreement with estimates of the peak intensity extracted from images of the focal area, obtained at reduced laser power. 
The original approach, based on the measurement of electron radiation spectra, could be extended further by using heavier particles, e.g.~protons. In this way, the applicability range broadens and the scheme could be able to gauge intensities in the range of $10^{25}$ W/cm$^{2}$.
Furthermore, a recent theoretical analysis \cite{marklund-arx19} of angular distributions of radiation emitted by an ultra relativistic electron beam interacting with an intense laser pulse also identified features which can be used for precise determination of the intensity value.
Finally, ponderomotive scattering of relativistic electrons has been theoretically examined in view of a more general task of the focus characterization \cite{mackenroth-arx19}.
Although this scheme involves several additional parameters characterizeing the initial and the scattered electron beams, it has also been shown potentially capable to measure the peak value of intensity in a laser focus.
These four schemes utilize essentially different physical effects, assume different interaction scenarios and therefore can be considered as independent and complimentary approaches to {\em in situ} diagnostics of ultrahigh laser intensities.

In this paper, we analyze the atomic diagnostics introduced in \cite{ciappina-pra19,ciappina-book19}.
Our purpose is to check the sensitivity of the method with respect to two essential factors present in experiment and in the theory.
Firstly, we check the sensitivity of our theoretical scheme to the model of tunneling ionization used to determine ionization rates which enter the system of rate equations for the populations of ionic charge states.
Secondly, we examine the focal volume effect on the distributions in ionic charge and check to which extent this volume effect smears out the sharp off-set used as an indicator of the peak laser intensity.
Our results show that at ultrahigh laser intensities of interest, the barrier suppression affects the charge distributions and can reduce the accuracy of the measurement.
We show that its influence minimizes for ionization of He- and H-like ions, and therefore the atomic specie should be selected such to reach the complete ionization in the central part of the laser focus.
Simultaneously, we show that the sharp dependance of the ion yield on the peak intensity value survives the focal averaging.
These results allow us to classify the suggested atomic diagnostics as a quantitatively reliable scheme which can be realized in practice.

The paper is organized as follows.
In the next section we discuss the interplay between the tunneling and the barrier suppression ionization (BSI) regimes. 
Section 3 presents an analysis of the focal averaging effect and, finally, the last section contains conclusions.
Atomic units are used throughout unless otherwise stated.

\section{Barrier-suppresion vs tunneling ionization}

In order to calculate the populations of different charge states in a strong laser field, we solve a system of rate equations (for more details see~\cite{ciappina-pra19}), using tunneling expressions for sequential ionization rates known as Perelomov-Popov-Terentiev (PPT) formulas \cite{ppt-jetp66a,pp-jetp67,popov-usp04,poprz-jpb14}.
Within the PPT theory, the time-dependent tunneling rate for a bound state with ionization potential $I_p$, orbital and magnetic quantum numbers $l$ and $m$, and residual charge $Z$ ($Z=1$ for ionization of neutral atoms) is given by 
\beq
w_{\rm TI}(\nu,l,m;t)=C_{\nu l}^2B_{lm}I_pF^{1+|m|-2\nu}(t)\exp\bigg\{-\frac{2}{3F(t)}\bigg\}~.
\label{w-lm}
\eeq
Here
\beq
\nu=\frac{z}{\sqrt{2I_p}}~,
\label{nu}
\eeq
is the effective principal quantum number and the coefficients $B_{lm}$ and $C_{\nu l}$ are 
\beq
B_{lm}=\frac{(2l+1)(l+|m|)!}{2^{2|m|}|m|!(l-|m|)!}~,~~~
C_{\nu l}^2=\frac{2^{2\nu-2}}{\nu\Gamma(\nu+l+1)\Gamma(\nu-l)}~.
\label{BC}
\eeq
Finally, the time-dependent reduced field $F(t)$ defined as:
\beq
F(t)=\frac{\sqrt{{\bf E}_L^2(t)}}{(2I_p)^{3/2}}~,~~~F\equiv {\rm max}\; F(t)=\frac{E_0}{(2I_p)^{3/2}}~,
\label{Ft}
\eeq
where ${\bf E}_L(t)$ is the laser electric field with an amplitude value $E_0$.

The choice of quasistatic rates is well justified by small values of the Keldysh parameter $\gamma=\sqrt{2I_p}\omega/E_0$ \cite{keldysh,popov-usp04,poprz-jpb14}, which fall into the interval $\gamma\simeq 10^{-2}\div 10^{-3}$ for ionization potentials $I_p\simeq 2\div 40$ keV, laser intensities $10^{20}\div 10^{24}{\rm W/cm}^2$ and laser frequencies $\omega\approx 0.05$ a.u., i.e. parameters considered in \cite{ciappina-pra19}. 
The tunneling ionization rates are asymptotically exact in the limit $F\to 0$ and remain quantitatively accurate in the domain $F\ll 1$ \cite{popov-usp04,poprz-jpb14}. 

At the same time, applicability of the tunneling approximation for quasistatic ionization rates can be considerably restricted from the side of high field strengths. 
Practically, for ground states of atoms and positive ions, the tunneling ionization rate (\ref{w-lm}) deviates from the respective exact value by $\sim 10\%$ at $F\approx 0.02...0.03$.
With a further growing of $F$ the tunneling formula quickly become quantitatively inaccurate.
The physical reason for that is in the suppression of the potential barrier the electron tunnels through.
This suppression leads to a violation of the WKB applicability conditions resulting in a considerable disagreement between the PPT and the exact ionization rates. 
In Ref.\cite{ciappina-pra19}, a criterion of fast ionization was formulated assuming that an initially populated atomic level is fully exhausted in a single optical cycle. 
This takes place at $F\simeq F_*\approx 0.05$ (see Eqs.~(13) and (14) there) and leads to the following empirical rule: an atomic level with a given ionization potential can hardly survive fields $F>F_*$.
Therefore if the tunneling rate (\ref{w-lm}) remains applicable up to $F\sim F_*$ the scheme of \cite{ciappina-pra19} is expected to be quantitatively reliable, otherwise effects of barrier suppression have to be included.

Following the 1D model of a hydrogen-like ion and neglecting the bound state Stark shift we may estimate the barrier-suppression (BS) field as $E_{\rm BS}\approx I_p^2/4Z$ which gives for the corresponding reduced field
\beq
F_{\rm BS}=\frac{E_{\rm BS}}{(2I_p)^{3/2}}\approx\frac{1}{16\nu}~.
\label{FBS}
\eeq
For the ground state of hydrogen, $F_{\rm BS}=1/16\approx 0.063>F_*$ while for ${\rm Kr}^{29+}$ with $I_p=3381$ eV and $\nu=1.84$, $F_{\rm BS}\approx 0.033<F_*$.
This means that, at intensities when fast ionization takes place, the actual ionization rates can in general considerably differ from their tunneling asymptotics.

This deviation can be accounted for numerically or by several approximate analytic methods (for a review of this long-standing problem, see e.g.~Chapter 3 of \cite{popov-usp04}, recent publications \cite{kost-pra18,kost-arx19} and references there).
Here we use an empiric formula proposed by Tong and Lin \cite{lin-jpb05}, which has been shown quite accurate in the intermediate domain $F\simeq F_{\rm BS}$.
According to \cite{lin-jpb05} the BS correction reduces to an additional factor in the rate:
\beq
w_{\rm BSI}(\nu,l,m;t)=w_{\rm TI}(\nu,l,m;t)\exp\bigg(-2\alpha\nu^2F(t)\bigg),
\label{WBS}
\eeq
where the fitting numerical coefficient $\alpha$ is chosen in the interval $6...9$ \cite{lin-jpb05}.
Depending on the values of $\nu$ and $\alpha$, the factor $\exp\bigg(-2\alpha\nu^2F\bigg)$ can vary from $\sim 2\div 3$ to $\simeq 10$, at $F\simeq F_{\rm BS}$.
Thus, the influence of the BS effect on the value ${\cal I}_{\rm s}$ of saturation intensity (i.e., intensity at which the given ionic state is generated with the probability close to unity, see a precise definition below) could be numerically significant.
Taking into account that no accurate analytic theory of BSI exists and the presently known models gain quantitatively different predictions (see, e.g. comparisons in \cite{kost-pra18}), there is a danger that the numerically found values of intensity ${\cal I}_{\rm s}$, may appear model-dependent.

\begin{figure}
\begin{center}
\includegraphics[height=8cm]{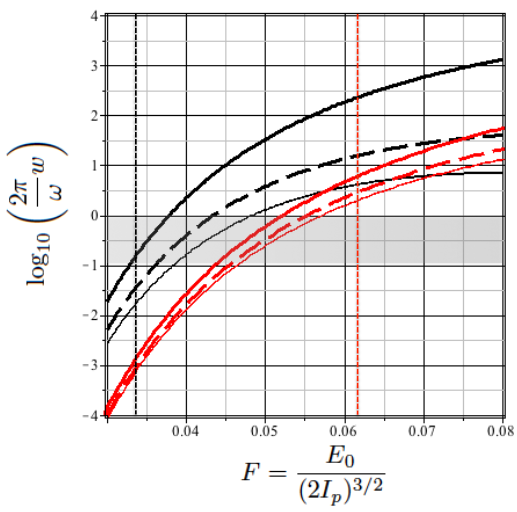}
\end{center}
\caption{(color online) Ionization probabilities $W=2\pi w/\omega$ per single cycle of a laser with wavelength $1\mu$m versus the peak reduced field $F$, calculated for ionization of ${\rm Kr}^{29+}$ with the electronic configuration $1s^22s^22p^3$ and ionization potential $I_p=3381$ eV (black lines) and of $1s^1$ ${\rm Ar}^{17+}$ with $I_p=4407$eV (red curves). Thick solid lines show the tunneling probability obtained from (\ref{w-lm}), dashed lines -- the BSI probability determined by (\ref{WBS}) with $\alpha=6$ and thin solid lines -- that with $\alpha=9$. The chosen interval of $F=0.03...0.08$ corresponds to the intensity interval $4.85\cdot 10^{20}...3.45\cdot 10^{21}{\rm W/cm}^2$ for Kr and $1.07\cdot 10^{21}...7.64\cdot 10^{21}{\rm W/cm}^2$ for Ar. Vertical dotted lines indicate the values of BS field (\ref{FBS}) equal to $0.033$ and $0.062$ correspondingly. A gray rectangular area shows the interval where the total probability per cycle changes from 0.1 (lower boundary) to 1.0 (upper boundary). Fast ionization occurs predominantly when the reduced field value falls into this interval.}
\label{Fig1}
\end{figure}

Curves of Fig.~1, where the total probability $W$ of ionization per laser cycle is shown versus the peak reduced laser field $F$, allow to qualitatively estimate the significance of the BS factor in (\ref{WBS}).
The main message of Fig.~1 is the relative insignificance of this factor for atomic states with $\nu\simeq 1$.
For ions of rare gases this corresponds to ionization of their $1s$ shell, i.e. of He- and H-like ions.
In this case the value of (\ref{FBS}) $F_{\rm BS}\approx 0.06>F_*\approx 0.05$ so that, except of extremely short laser pulses which can not be realized with $\mu{\rm m}$-scale wavelengths, such shells have enough time to be ionized in the regime of tunneling before the filed amplitude enters the BS domain.
Consequently, the BS correction is only expected to slightly enhance the saturation intensity ${\cal I}_{\rm s}$.
In contrast, the ionization probabilities for outer shells with $\nu\approx 2\div 3$ demonstrate a stronger sensitivity to the BS regime, because the bound state lies higher and therefore $F_{\rm BS}<F_*$.
For such levels, the tunneling ionization process may happen too slow to fully strip the level out before the barrier is suppressed by the growing laser field.
Thus, for $2s,~2p$ and higher shells the value of ${\cal I}_{\rm s}$ can be considerably enhanced by the BS effect. 

In order to quantify the effect of BSI on the value of ${\cal I}_{\rm s}$ we numerically solve the system of rate equations with the tunneling (\ref{w-lm}) and BS (\ref{WBS}) ionization rates for  argon and krypton in the intensity interval $10^{19}\div 10^{22}{\rm W/cm}^2$.
The laser pulse shape and duration are the same as in \cite{ciappina-pra19}:
\beq
E(t)=E_0\cos(\omega t)\sin^{2}\left(\frac{\omega t}{2 N}\right)~.
\label{Et}
\eeq
In our simulations we use $\omega=0.0455$ a.u. and $N=10$, that correspond to a laser wavelength $\lambda=1$ $\mu$m and a total pulse length $T\sim33$ fs, respectively.
The systems of rate equations for populations $0\le c_n\le 1$ of ions $A^{n+}$ are taken from \cite{ciappina-pra19}.

\begin{figure}
\begin{center}
\leftline{\includegraphics[height=10cm]{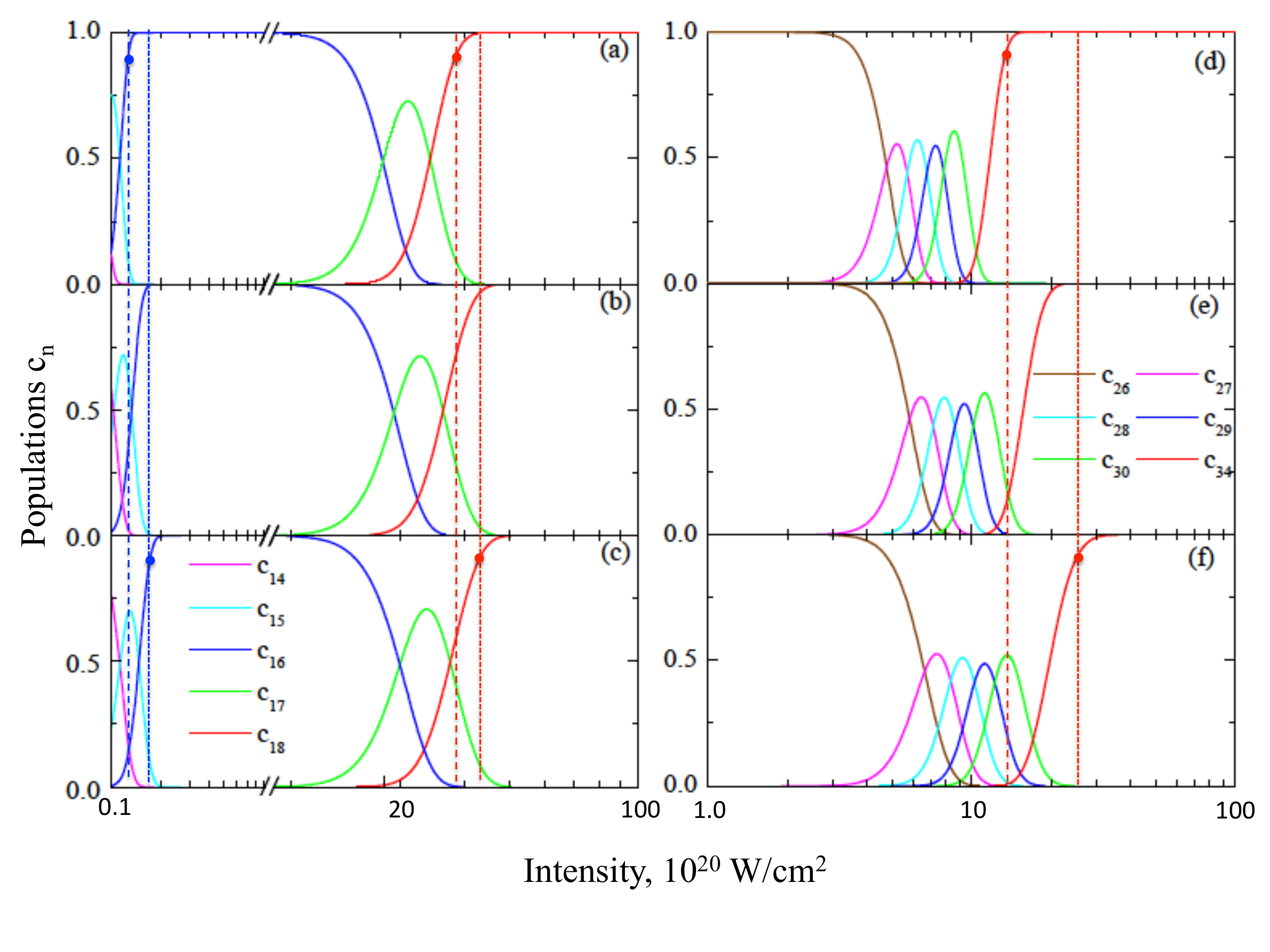}}
\end{center}
\caption{(color online) Populations $c_n$ of ionic states at the end of the laser pulse (\ref{Et}) calculated by solving numerically the system of rate equations for argon (left column) and krypton (right column) using the tunneling (first row) and BS ionization rates with $\alpha=6.0$ (second row) and $\alpha=9.0$ (third row). The initial populations are set $c_{14}(0)=1.0$, $c_{15}(0)=...c_{18}(0)=0$ for argon and $c_{26}(0)=1.0$, $c_{27}(0)=...c_{36}(0)=0$ for krypton. Vertical lines indicate the saturation intensities ${\cal I}_{\rm s}$ defined such that the corresponding $c_n=0.9$, for ${\rm Ar}^{16+}$ (blue lines), ${\rm Ar}^{18+}$ (red lines) and ${\rm Kr}^{34+}$ (red lines). Dashed and dotted lines indicate the values of ${\cal I}_{\rm s}$ for the tunneling and BS ionization mechanisms correspondingly.}
\label{Fig2}
\end{figure}

Figure 2 shows the intensity-dependent populations at the end of the laser pulse calculated for ionization cascades developing from ${\rm Ar}^{14+}$ and ${\rm Kr}^{26+}$.
These initial states were chosen different from those for neutral atoms to greatly reduce the number of rate equations in the systems.
Practical independence of ionization cascades for high charge states on the initial condition was demonstrated in \cite{ciappina-pra19}.
The value of saturation intensity ${\cal I}_{{\rm s}n}$ is defined such that $c_n({\cal I}_{{\rm s}n})=0.9$.
The BS effect is clearly seen.
It mostly reduces to the overall shift of the distributions toward higher intensities.
As suggested by the analysis above, for ionization of He- and H-like ions ${\rm Ar}^{16+}$ and ${\rm Ar}^{17+}$ this shift appears smaller than for other states where $\nu\approx 2$.
To quantify the BS effect we introduce a relative shift of the saturation intensity
\beq
\delta({\rm A}^{n+})=\frac{{\cal I}_{{\rm s}n}(\alpha=9)-{\cal I}_{{\rm s}n}(\alpha=0)}{{\cal I}_{{\rm s}n}(\alpha=9)+{\cal I}_{{\rm s}n}(\alpha=0)}
\label{delta}
\eeq
For parameters of Fig.2, Eq.(\ref{delta}) gives $\delta({\rm Ar}^{18+})\approx 0.07$, $\delta({\rm Ar}^{16+})\approx 0.16$ and $\delta({\rm Kr}^{34+})\approx 0.30$.
Moreover, for ${\rm Ar}^{16+}$ and ${\rm Kr}^{34+}$ the shift in the saturation intensity $\Delta{\cal I}_{\rm s}$ is comparable to the value of intensity interval where the number of ions of given charge grows by approximately one order in magnitude, while for ${\rm Ar}^{18+}$ this value is approximately 3 times smaller.
These observations are in favor of using $1s$ states for determination of the peak intensity value.

\section{Focal volume effect}

The ionic signal collected by a time-of-flight (TOF) detector presents a distribution in charge states averaged over the whole focal volume or its significant part.
This space averaging affects both the distribution shape and the intensity dependence of the yield for a given charge state.
Here we focus on the latter and calculate the number of ions produced in the laser focus using results obtained in Ref.\cite{ciappina-pra19} and in Section 2 above.
Taking into account that the accuracy of the method appears the highest for ionization of He- and H-like ions, we restrict the analysis by this case and use the pulse form (\ref{Et}) of the previous section.
Ionic states populations $c_n({\cal I})$ at the end of the laser pulse are given by plots of Fig.2 (see also Figs.4-7 in \cite{ciappina-pra19}).
Using these data, one can calculate the total number of ions produced in the focus by replacing ${\cal I}\to{\cal I}(\vecr)$, choosing a model for the focal intensity distribution ${\cal I}(\vecr)$ and integrating over the focal volume:
\beq
N({\rm A}^{n+})=n_0\int c_n({\cal I}(\vecr))d^3r~.
\label{N}
\eeq
Below we assume the Gaussian beam
\beq
{\cal I}(\rho,z)=\frac{{\cal I}_0w_0^2}{w^2(z)}\exp\bigg(-\frac{2\rho^2}{w^2(z)}\bigg)~,~~~w(z)=w_0\sqrt{1+\frac{z^2}{z_R^2}}~,~~~z_R=\frac{\pi w_0^2}{\lambda}
\label{gauss}
\eeq
with $w_0$ and $z_R$ being the beam waist and Rayleigh length correspondingly, and the wavelength is taken $\lambda\simeq 1\mu{\rm m}$.

Using the numerically found populations $c_n({\cal I})$ we calculated the number of ions ${\rm Ar}^{17+}$ and ${\rm Ar}^{18+}$ from (\ref{N}) as functions of the peak intensity value ${\cal I}_0$ for the case when the tunneling theory of ionization applies $(\alpha=0)$.
The initial gas concentration is taken $n_0=2\cdot 10^{12}{\rm cm}^{-3}$ corresponding to the pressure $3\cdot 10^{-5}$Torr at room temperature.
At these parameters, the number of atoms in the central part of the focus $N_0\approx 3\cdot 10^3$.
Results shown on Fig.3 demonstrate a steep growth in the total number of ions by two orders in magnitude when the peak intensity increases by approximately factor of two, between $3\cdot 10^{21}$ and $5\cdot 10^{21}{\rm W/cm}^2$ for ${\rm Ar}^{18+}$.
Because of the relatively small difference in ionization potentials of the $1s^2$ and $1s$ states of argon, the number ${\rm Ar}^{18+}$ ions can exceed that of ${\rm Ar}^{17+}$ because of the volume effect, that happens at ${\cal I}_0\approx 7\cdot 10^{21}{\rm W/cm}^2$.
Ion yields for ${\rm Xe}^{53+}$ and ${\rm Xe}^{54+}$ are shown on the inset.
They demonstrate a similar steep two orders of magnitude growth in the intensity interval $(1.5\div 2.0)\cdot 10^{24}{\rm W/cm}^2$.
The yields become equal at ${\cal I}_0\approx 6.5\cdot 10^{24}{\rm W/cm}^2$ (not shown on the plot).
This steep increase in the number of ions results from the simultaneous action of the focal volume effect and of the increase of the coefficient $c_n$ as shown on Fig.2. 

\begin{figure}
\begin{center}
\leftline{\includegraphics[height=8cm]{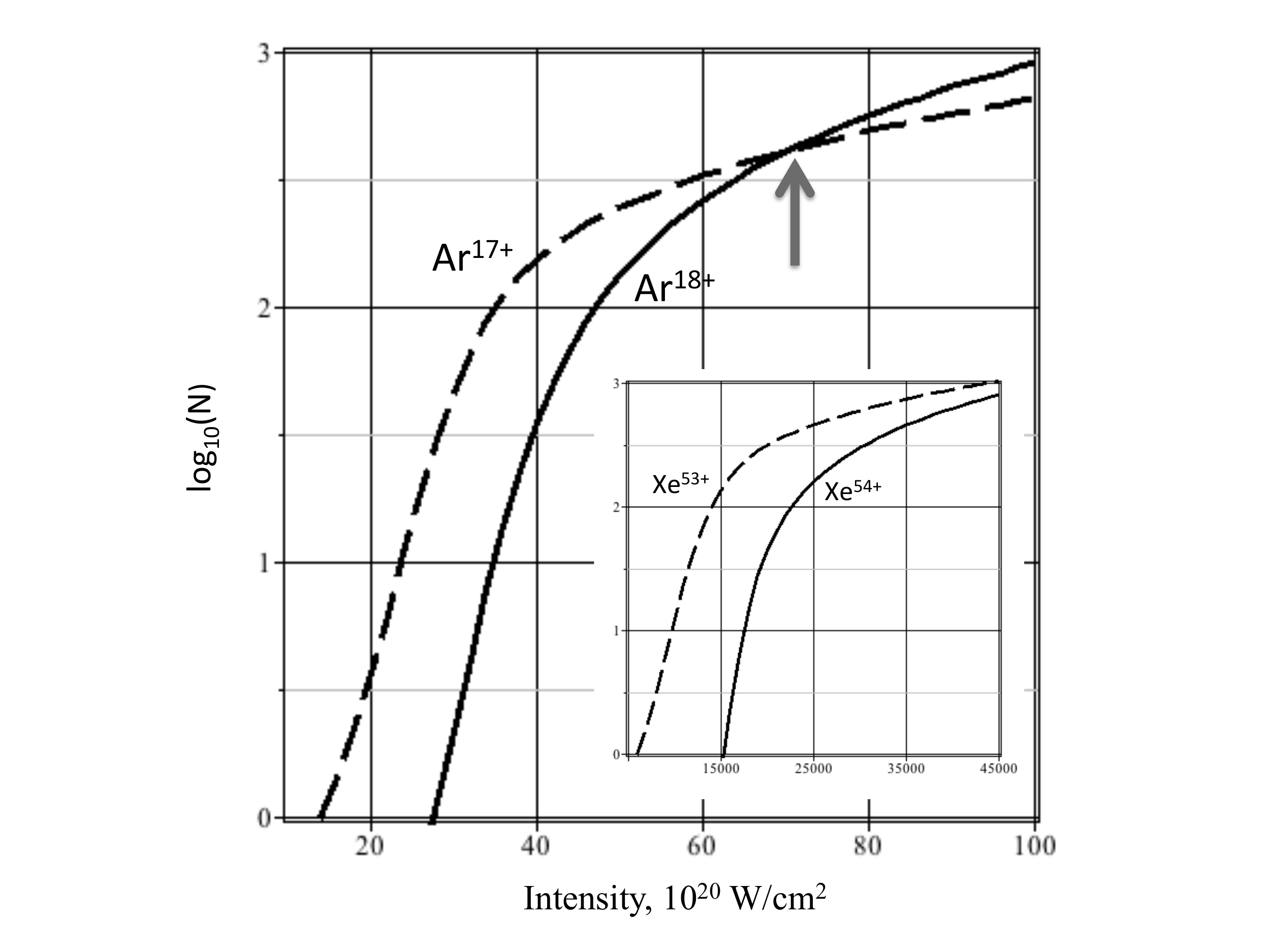}}
\end{center}
\caption{Total number of ions $N({\rm A}^{n+})$ produced in a Gaussian laser focus with waist $w_0=3\mu{\rm m}$ and concentration of neutral atoms $n_0=2\cdot 10^{12}{\rm cm}^{-3}$. The main plot shows the intensity dependence of $N$ for ${\rm Ar}^{17+}$ and ${\rm Ar}^{18+}$; analogous dependences for ${\rm Xe}^{53+}$ and ${\rm Xe}^{54+}$ are shown on the inset. A gray arrow indicates the intensity ${\cal I}_0\approx 7\cdot 10^{21}{\rm W/cm}^2$ where $N({\rm Ar}^{17+})=N({\rm Ar}^{18+})$.}
\label{Fig3}
\end{figure}

Both the initial steep part of the ion yield curve and the intersection point can be used for determination of the intensity value. 
The main advantage of the first case is that the steep part of the yield curve only weakly depends on the focal spot shape. 
This becomes clear when we note that for intensities only slightly exceeding the ionization threshold, the ion signal comes from a small vicinity of the focus center (i.e. the point where the intensity reaches its peak value) where the intensity distribution can be described by a parabolic shape.
From the other side, if the effectiveness of a time of flight detector, which depends in particular on the collection angle, is not precisely known, a direct correspondence between the number of ions shown on Fig.3 and the number of counts cannot be established.
Then the intensity value can only be located within the interval corresponding to the steep slope of the yield.
This imposes the lower limit on the measurement uncertainly equal to $\Delta {\cal I}_0\approx 2\cdot 10^{21}{\rm W/cm}^2$ for argon and $\Delta {\cal I}_0\approx 0.5\cdot 10^{24}{\rm W/cm}^2$ for xenon which gives a $50\%$ and $30\%$ accuracy correspondingly.
The second method based on the determination of the intersection point, $N(A^{n+})=N(A^{n+1})$, can potentially provide a higher accuracy.
However the intersection of the yields happens when the laser intensity exceeds its threshold value already in the significant part of the focal spot.
As a result, the position of this point can considerably depend on the focal shape introducing an additional source of uncertainty.

\section{Conclusions}

In conclusion, we presented a numerical analysis on the scheme for experimental determination of the peak value of intensity in a focus on a multy-petawatt power laser beam, based on tunneling ionization of multielectron atoms \cite{ciappina-pra19}.
Our analysis shows that:
\begin{itemize}
\item {The saturation intensity value ${\cal I}_s$ appears sensitive to the barrier-suppresion correction to the ionization rate which makes the whole algorithm of  intensity extraction dependent on the model of BS ionization.
This dependence is however not crucially strong and appears minimal for ionization of He- and H-like ions with effective principal quantum numbers $\nu\approx 1$.
Thus such ions should be used as intensity markers.}
\item {The highly nonlinear dependence of the ion yield on intensity at ${\cal I}\approx {\cal I}_{s}$ survives the focal averaging procedure and the steep part of the yield curve can be used to measure intensity with accuracy $30\div 50\%$.}
\item{A simultaneous determination of the intersection point where $N(A^{n+})=N(A^{(n+1)+})$ can provide some information on the intensity distribution in the focal spot.}
\end{itemize}
These results can be used to design experiments on characterization of extremely intense electromagnetic pulses delivered by the future multi-PW and exawatt laser sources.

\section*{Acknowledgments}
Authors acknowledge useful discussions with S.V. Bulanov, G. Korn, D. Kumar and S. Weber.
SVP acknowledges financial support of the Russian Foundation for Basic Research via Grant No.19-02-00643 and of the Ministry of Education
and Science of the Russian Federation via Grant No.3.1659.2017/4.6. This work was also supported by the project Advanced research using high intensity laser produced photons and particles (No. CZ.02.1.01/0.0/0.0/16\_019/0000789) from European Regional Development Fund (ADONIS).


\section*{References}
\begin{thebibliography}{9}

\bibitem{APRI} Sung J H, Lee H W, Yoo J Y \etal 2017 {\em Opt. Lett.} {\bf 42} 2058

\bibitem{CAEP} Zeng X \etal 2017 {\em Opt. Lett.} {\bf 42} 2014

\bibitem{SULF} Gan Z \etal 2017 {\em Opt. Exp.} {\bf 25} 5169; Li W \etal {\em Opt. Lett.} {\bf 43} 5681

\bibitem{ELI} Chambaret J-P, Chekhlov O, Cheriaux G \etal 2010 Extreme light infrastructure: laser architecture and major challenges, in ``Solid State Lasers and Amplifiers IV, and High-Power Lasers'' {\bf 7721} 77211D

\bibitem{P3-ELI} Weber S \etal 2017 {\em Mat. Radiat. Extr.} {\bf 2} 149 

\bibitem{XCELS} Bashinov A V, Gonoskov A A, Kim A V, Mourou G and Sergeev  A M 2014 {\em  Eur. Phys. J.  Spec. Top.} {\bf 223} 1105.

\bibitem{bulanov-rmp09} Mourou G, Tajima T and S.V. Bulanov S V 2009 \rmp {\bf 78} 309

\bibitem{dipiazza-rmp12} Di Piazza A, M\"{u}ller C, Hatsagortsyan  C Z and Keitel C H  2012 \rmp {\bf 84} 1177

\bibitem{fedotov-cp15} Narozhny N B and Fedotov A M 2015 {\em Contemp. Phys.} {\bf 56} 249

\bibitem{ciappina-pra19} Ciappina M F \etal 2019 \pra {\bf 99} 043405

\bibitem{ciappina-book19} Ciappina M F \etal 2019 in {\em Progress in Ultrafast Intense Laser Science XV} ed. by D. Charalambidis and K. Yamanouchi, Springer, in press
\bibitem{walker-pra01} Chowdhury E A, Barty C P J and Walker B C 2001 \pra {\bf 63} 042712

\bibitem{walker-josa03} Chowdhury E A and Walker B C 2003 {\em J. Opt. Soc. Am. B} {\bf 20} 109

\bibitem{yamakawa-pra03} Yamakawa K, Akahane Y, Fukuda Y, Aoyama M, Inoue N and Ueda H 2003 \pra {\bf 68} 065403

\bibitem{yamakawa-jmo03} Yamakawa K, Akahane Y, Fukuda Y, Aoyama M, Ma J, Inoue N, Ueda H and Kiriyama H 2003 \jmo {\bf 50} 2515

\bibitem{ppt-jetp66a} Perelomov A M, Popov V S and Terentev M V 1966 {\em Zh. Eksp. Teor. Fiz.} {\bf 50} 1393 [1966 {\em Sov. Phys. JETP} {\bf 23} 924 (Engl. transl.)]

\bibitem{pp-jetp67} Perelomov A M and Popov V S 1967 {\em Zh. Eksp. Teor. Fiz.} {\bf 52} 514 [1967 {\em Sov. Phys. JETP} {\bf 25} 336 (Engl. transl.)]

\bibitem{popov-usp04} Popov V S 2004 {\em Usp. Fiz. Nauk} {\bf 147} 921 [2004 {\em Phys. Usp.} {\bf 47} 855 (Engl. transl.)]

\bibitem{poprz-jpb14} Popruzhenko S V 2014 \jpb {\bf 47} 204001 

\bibitem{he-oe19} He C Z, Longman A, P\'erez-Hern\'andez J A, de Marco M, Salgado C, Zeraouli G, Gatti G, Roso L, Fedosejevs R and Hill W T 2019 {\em Opt. Exp.} {\bf 27} 30020

\bibitem{marklund-arx19} Blackburn TG, Gerstmayr E, Mangles S P D, Marklund M 2019 arXiv:1911.02349

\bibitem{mackenroth-arx19} Mackenroth F, Holkundkar A R, Schlenvoigt H P 2019 arXiv:1909.05008

\bibitem{keldysh} Keldysh L V 1964 \jetp \textbf{47} 1945 [1965 {\em Sov. Phys. JETP} \textbf{20} 1307 (Engl. transl.)] 

\bibitem{kost-pra18} Kostyukov I Y and Golovanov A A 2018 \pra {\bf 98} 043407

\bibitem{kost-arx19} Kostyukov I Y and Golovanov A A 2019 arXiv:1906.01358

\bibitem{lin-jpb05} Tong X M and Lin C D 2005 \jpb {\bf 38} 2593













\end{thebibliography}
\end{document}